\documentclass[12pt, letterpaper]{article}

\usepackage{graphicx}
\usepackage[breaklinks, ps2pdf]{hyperref}
\usepackage{threeparttable}
\usepackage{natbib}
\renewcommand{\cite}{\citep}
\usepackage{amssymb}
\renewcommand{\phi}{\varphi}
\renewcommand{\epsilon}{\varepsilon}
\begin{document}


\title{Effects of missing data in social networks \thanks{The author
thanks Peter Dodds, Nobuyuki Hanaki, Alexander Peterhansl, Duncan
Watts and Harrison White for careful reading of the manuscript and
many useful comments, and Mark Newman for kindly providing the Los
Alamos E-print Archive collaboration data.}  }

\author{ \protect G. Kossinets \thanks{ Institute for Social and
Economic Research and Policy, Columbia University, 420 W. 118th St.,
2nd floor, Mail Code 3355, New York, NY 10027.  Tel.: +1-212-854-0367,
fax: +1-212-854-7998, email: gk297@columbia.edu. } }

\date{ \today}
\maketitle 







\begin{abstract} 

We perform sensitivity analyses to assess the impact of missing data
on the structural properties of social networks.  The social network
is conceived of as being generated by a bipartite graph, in which
actors are linked together via multiple interaction contexts or
affiliations.  We discuss three principal missing data mechanisms:
network boundary specification (non-inclusion of actors or
affiliations), survey non-response, and censoring by vertex degree
(fixed choice design), examining their impact on the scientific
collaboration network from the Los Alamos E-print Archive as well as
random bipartite graphs.  The results show that network boundary
specification and fixed choice designs can dramatically alter
estimates of network-level statistics.  The observed clustering and
assortativity coefficients are overestimated via omission of
interaction contexts (affiliations) or fixed choice of affiliations,
and underestimated via actor non-response, which results in inflated
measurement error.  We also find that social networks with multiple
interaction contexts have certain surprising properties due to the
presence of overlapping cliques.  In particular, assortativity by
degree does not necessarily improve network robustness to random
omission of nodes as predicted by current theory.  \\


(Model Evaluation and Testing).

\end{abstract}

\newpage



\section{Introduction}
\label{Intro}
Social network data is often incomplete, which means that some actors
or links are missing from the dataset.  In a normal social setting,
much of the incompleteness arises from the following main sources: the
so-called Boundary Specification Problem \cite{LaMa83}; respondent
inaccuracy \cite{BeKi84}; non-response in network surveys \cite{Ru93};
or may be inadvertently introduced via study design.  Although missing
data is abundant in empirical studies, little research has been
conducted on the possible effect of missing links or nodes on the
measurable properties of networks at large.
In particular, a revision of the original work done primarily in the
1970-80s \cite{HoLe73, LaMa83, BeKi84} seems necessary in the light of
recent advances that have brought new classes of networks to the
attention of the interdisciplinary research community \cite{AmSc00,
BaAl99, NeSt01, St01, WaSt98, Wa99}.

Let us start with a few examples from the literature to illustrate
different incarnations of missing data in network research.  The
boundary specification problem \cite{LaMa83} refers to the task of
specifying inclusion rules for actors or relations in a network study.
Researchers who study intraorganizational networks typically ignore
numerous ties that lead outside an organization, reasoning that these
ties are irrelevant to the tasks and operations that the organization
performs.  A classical account is the Bank Wiring Room study
\cite{RoDi39}, which focused on 14 men in the switchboard production
section of an electric plant.  The sociometric data obtained in that
study have been analyzed extensively~\cite{Ho50, WhBo76} but the
effect of interactions outside the wiring room on the workers'
behavior and performance at work is unknown and hardly feasible to
estimate.

In a recent study of romantic relationships in a large urban high
school \cite{BeMo02}, more than one half of relationships reported in
the period of 18 months were with persons who did not attend the
school.  The network appears to have a large connected component
linking together about one half of romantically involved pupils.  The
authors proposed an elegant explanation for the observed structure in
terms of a micro-social norm governing the pair-formation process.
However, by focusing solely on the in-school network, the authors
implicitly assumed that the remaining 60\% of relationships had little
effect on social dynamics within the school community.  Such a large
fraction of outside nominations makes one wonder if homogeneity of
dating norms within the school may be affected by student liaisons
with the larger community in which the school is embedded.


The boundary specification problem may be avoided to a certain extent
if the community is isolated from the rest of the world as e.g. in
Sampson's monastery \cite{Sa69}.  By and large, however, network
closure is an artifact of research design, i.e. the result of
arbitrary definition of network boundaries.  When choosing inclusion
rules for a network study, a researcher is effectively drawing a
non-probability sample from all possible networks of its kind
\cite{LaMa83}.  As a result, it is almost impossible to estimate
the error introduced into network data via study design.  Dynamic
changes in the network (waxing and waning relationships or activation
of latent ties) only exacerbate the problem.

The problem of {\em informant inaccuracy} has enjoyed more close
attention in the last decades \cite{BeKi84, Ma90} and basically
represents the case where respondents take their perception of a
social relation for the relation itself.  As a consequence, network
data collected by interviewing or administering a network instrument
may reflect the cognitive network rather than the actual interaction
pattern.  In particular, it has been found that the discrepancy
between cognitive and real network in recall data depends on time in a
curiously non-linear fashion \cite{BeKi84}.  Some ways of alleviating
this problem have been proposed, and good network instruments help
minimize this kind of bias.  At times, however, the cognitive network
might be exactly what the researcher is looking for (e.g., in
marketing applications, etc.).  On the other hand, many social
transactions such as electronic mail may be registered directly and
data thus obtained does not contain a significant idiosyncratic
component.  In this paper we do not explicitly model the effect of
informant inaccuracy, assuming that either it is consistent with the
research framework, or that the network in question was reconstructed
from reliable electronic, historical or survey data.

An important problem in network survey research is that of {\em survey
non-response}.  In a standard sampling situation such as drawing a
representative sample from some population, special techniques are
available to correct parameter estimates for imperfect response rates
\cite{LiRu02}.  Unfortunately, no such definitive treatment is
available for social network analysis, although effects of
non-response on some network properties have been described previously
\cite{StRi92, Ru93}.  We generally follow this exploratory
line of research in that we discuss how network structure is affected
by different non-response scenarios and propose some ways to
ameliorate the problem.

Compound missing data mechanisms may be encountered as well; a good
example is forensic network research.  Besides fuzzy boundaries,
criminal networks are characterized by presence of unknown actors,
actors with false identities, and hidden or dormant ties \cite{Sp91}.
Network analysis practitioners have noticed that minor changes in
graph structure (addition or deletion of vertices or links) can have a
dramatic effect on network properties as a whole, especially on
individual-level indices \cite{Kr02}.  The extent of the distortion
depends on the nature of group structure itself as well as on data
collection and analysis procedures \cite{HoLe73}.  However, the
sensitivity of many graph-theoretic measures to missing data,
especially of those introduced recently, has not been assessed
numerically.  Not all graph-theoretic indices are applicable to
criminal network research from an epistemological point of
view,\footnote{\citet{Sp91} notes that ``fuzzy
boundaries render precise global measures (such as radius, diameter,
even density) almost meaningless'' and suggests that {\em betweenness
centrality} is probably the most useful measure for criminal
networks.} and yet fewer may be reliable enough with respect to
missing data.

Social network data may as well be biased as a result of study design.
In this paper we analyze the so-called {\em fixed choice effect}
\cite{HoLe73}.  Consider a friendship network in which actors have
anywhere between 1 and 10 friends each.  Often network researchers ask
respondents to make nominations only up to some fixed number.  Suppose
that we asked our participants to write down up to three best friends
of theirs.  How is the network constructed in that particular way
different from the ``true'' friendship network?  Does the effect
depend on structural properties of the friendship graph?  These are
some of the questions that we aim to answer.

This paper aims to fill the methodological vacuum around the problem
of missing data in social network analysis.  One approach to deal with
it is to develop analytic techniques that capture global statistical
tendencies and do not depend on individual interactions \cite{RaHo61}.
A complmentary strategy is to develop remedial techniques that
minimize the effect of missing data \cite{HoLe73}.  Although we do not
offer a definitive statistical treatment in this paper, we conduct
exploratory analyses and advocate the importance of further work in
this direction.\footnote{After this manuscript was completed we became
aware of another study with a similar approach that focused
exclusively on different network centrality measures \cite{CoVa03}.}
To explore the problem and outline possible solutions we use the
method of statistical simulation.  The general outline of our approach
is as follows: (1) take a real (large enough) social network or an
ensemble of random graphs and assume that network data is complete;
(2) remove a fraction of entities to simulate different sources of
error; and (3) measure network properties and compare to the ``true''
values (from the ``complete'' network). We quantify the uncertainty
caused by missing network data and assess sensitivity of graph-level
metrics such as average vertex degree, clustering coefficient
\cite{NeSt01}, degree correlation coefficient \cite{Ne02}, size and
mean path length in the largest connected component.

We illustrate the problem using the scientific collaboration graph
containing authors and papers from the Condensed Matter section of the
Los Alamos E-print Archive from 1995 through 1999~\cite{Ne01} and use
this example to develop a statistical argument for the general case of
social networks with multiple interaction contexts.  Owing to the
sheer size of the dataset, the numerical estimates have very narrow
confidence intervals.  The results are compared to the case of random
bipartite graphs.

The paper is organized as follows.  Section \ref{Sources} focuses on
the sources of missing or false data in social network research.  We
generalize the Boundary Specification Problem (BSP) for social
networks with multiple interaction contexts modeled as bipartite
graphs, in which actors are linked via multiple affiliations or
collaborations.  We discuss the issues of non-response and
non-reciprocation in social network studies as well as the degree
cutoff bias often introduced by questionnaire design.  Section
\ref{Data} describes relevant network statistics, datasets and
simulation algorithms that are used to investigate effects of missing
data on network properties.  Section \ref{Results} presents the
results, while Section \ref{Conclusions} summarizes the findings and
discusses a number of potential applications.


\section{Sources of missing data in social networks}
\label{Sources}


\subsection{The Boundary Specification Problem}
\label{BSP}

Network boundary specification which consists of defining rules for
inclusion of actors (and relations) in the network under
investigation, is a major epistemological problem in social network
research.  It was first addressed by \citet{LaMa83} who identified
three basic strategies in dealing with the problem.  Of course
multiple inclusion strategies are possible, as a logical combination
of those discussed here.

According to the {\em nominalist approach}, actors are included in the
network based on the formal definition of group membership (recall
examples in the beginning of the paper).  Detailed specifications can
factor in actors' attributes (all non-white first year students of a
college), relations (all respondents who reported being involved in a
romantic relationship), events (all individuals who attended a college
party), etc., whereby a conceptual framework is imposed by the analyst
and the network boundary becomes devoid of ontologically independent
status~\cite{LaMa83}.  The last example (event attendance) is
particularly error-prone and is best described as convenience
sampling, with non-generalizable results and all sorts of biases
operating including self-selection (e.g. people who attend an event
may be quite gregarious and therefore different from those who do
not attend).

One particular instance of the nominalist approach is positional
specification, most commonly defined as occupancy of a ranked position
in a formally constituted group.  Examples include a country's 100
best known politicians, or 500 top business firms
\cite[e.g.][]{DaMi99}.  This approach involves setting an arbitrarily
limiting scope in order to facilitate analysis
or due to data availability.  It is important to know whether network
data thus obtained is susceptible to data-specific and subjective
bias.

The {\em realist approach} (in the Marxist sense) lets actors
themselves define network boundaries. ``The network is treated as a
social fact only in that it is consciously experienced as such by the
actors composing it''~\cite{LaMa83}.  A particular example would be
recognized common membership status (students, etc.).  This approach
emphasizes the cognitive dimension over social interactions per se;
hence it may be more susceptible to informant inaccuracy effects.
Actors may disagree in their perception of social structure; they may
be attributing different weights to certain other actors, relationships
or types of relationships.  The correspondence between analytically
drawn boundaries and the ``collectively shared subjective awareness''
of these boundaries by the actors should be treated as an empirical
question rather than an assumption \cite{LaMa83}.

Finally, an {\em empiricist approach} aims to go beyond cognitive
experience of either the researcher or social actors and instead
focuses on measurable interactions.  The network boundary is defined
by recording who is interacting with whom in a certain context.  This
approach has not been feasible for large networks until recently, when
data on large-scale social interactions become readily available from
the records of email communication or virtual communities
\cite{EbMi02, GuDa02, HoEd02, NeFo02}. The empiricist approach
requires an operational specification of the interaction setting or
context, and then including all actors who interact within this
context.  The missing data mechanism associated with this approach is
the boundary specification problem for relations.

\begin{figure} 
\begin{center}
\includegraphics[width=0.25\textwidth]{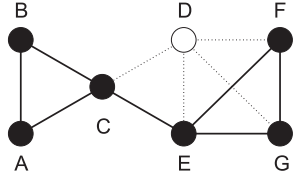}
\end{center}
\caption{Illustration of the Boundary Specification Problem.  Omission
of actors may lead to significant changes in network statistics.  In
the above example, as a result of exclusion of actor $D$, the mean
network degree $z$ went down 25\% from $3\frac{1}{7}$ to
$2\frac{1}{3}$. }
\label{fig:bspa}
\end{figure}


\subsection{The boundary specification problem for relations}

Since social networks are constructed from actors and relations
between actors, the boundary specification problem has two faces to
it.  In addition to defining a network boundary over the set of
actors, researchers make arbitrary decisions on which relations to
consider.  Often it is determined by the task at hand, e.g. a study of
the spread of HIV would perhaps include only two relations (sexual
contacts and needle sharing) without any loss of validity.  For other
interesting topics, such as collective movements or social contagion
processes, relevant network relations are not so easy to define.

Consequently, a researcher of social networks faces the question of
what types of links to include.  This problem is conceptually close to
the task commonly faced in the traditional social research focused on
individual attributes, that is, which variables should be analyzed.
Usually the research is informed by theory and aided by exploratory
numerical techniques (as in econometrics and finance).  Yet there is
no consistent theory of social interactions to guide network
research~\cite{Wh92}, which leaves us face-to-face with a non-trivial
epistemological problem.  Laumann et al. propose that key ties may be
omitted ``due to oversight or use of data that are merely convenient.
Such an error, because it distorts the overall configuration of actors
in a system, may render an entire analysis
meaningless''~\cite{LaMa83}.

We develop here a {\em multicontextual} approach based on actors'
participation in groups, events or activities.  The key idea is to
break down social ties to identifiable, discrete interactions.  As we
have illustrated, social actors belong to multiple affiliations,
attend various events, participate in different interaction contexts,
and every interaction may be important for the dynamics of the social
network in which actors are embedded~\cite{Br74, Wh92}.

The idea that people participate in multiple relations with one
another is certainly quite old \cite[cf.][]{Simmel}, so it
seems surprising that only a few studies have made use of multiple
interaction contexts in mathematical models of social networks.
\citet{WhBo76} demonstrated in 1976 that it is possible to efficiently
extract an image of social structure underlying multiple relations
defined for the same set of actors.  \citet{WaDo02}, based on the
results of \citet{TrMi69} as well as their own recent electronic
experiment \cite{DoMu03}, proposed that people use multiple relations in
order to solve the small world problem, i.e. to deliver a message to
an unknown target using only connections from within their egocentric
network.  In both studies, however, the number of actors is much
greater than the number of relations in which actors participate.
Perhaps this might be an artifact of study design when researchers
combine several relations in one group to prevent possible
misunderstanding on part of human subjects.  On the other hand, this
might be an indication that actors themselves group similar relations
into broader and therefore more robust classes of relations.
There may be several reasons for doing this: (1) relations may be
correlated, e.g. when one relation almost always implies another; (2)
people may (mis-)perceive and assign varying importance to relations
in an idiosyncratic fashion; (3) people may manipulate relations,
e.g. using personal ties to gain power in an organization.  In
general, it seems hardly possible to disentangle the manifold of
social interactions (group and dyadic, etc.) that make up social
fabric.  It is the joint network, made by juxtaposition of all
relevant kinds of ties between actors, that matters in dynamics of
processes based on social influence \cite{Wh92, WhBo76}.

\begin{figure} 
\begin{center}
$\begin{array}{c@{\hspace{0.7in}}c}
\includegraphics[width=0.3\textwidth]{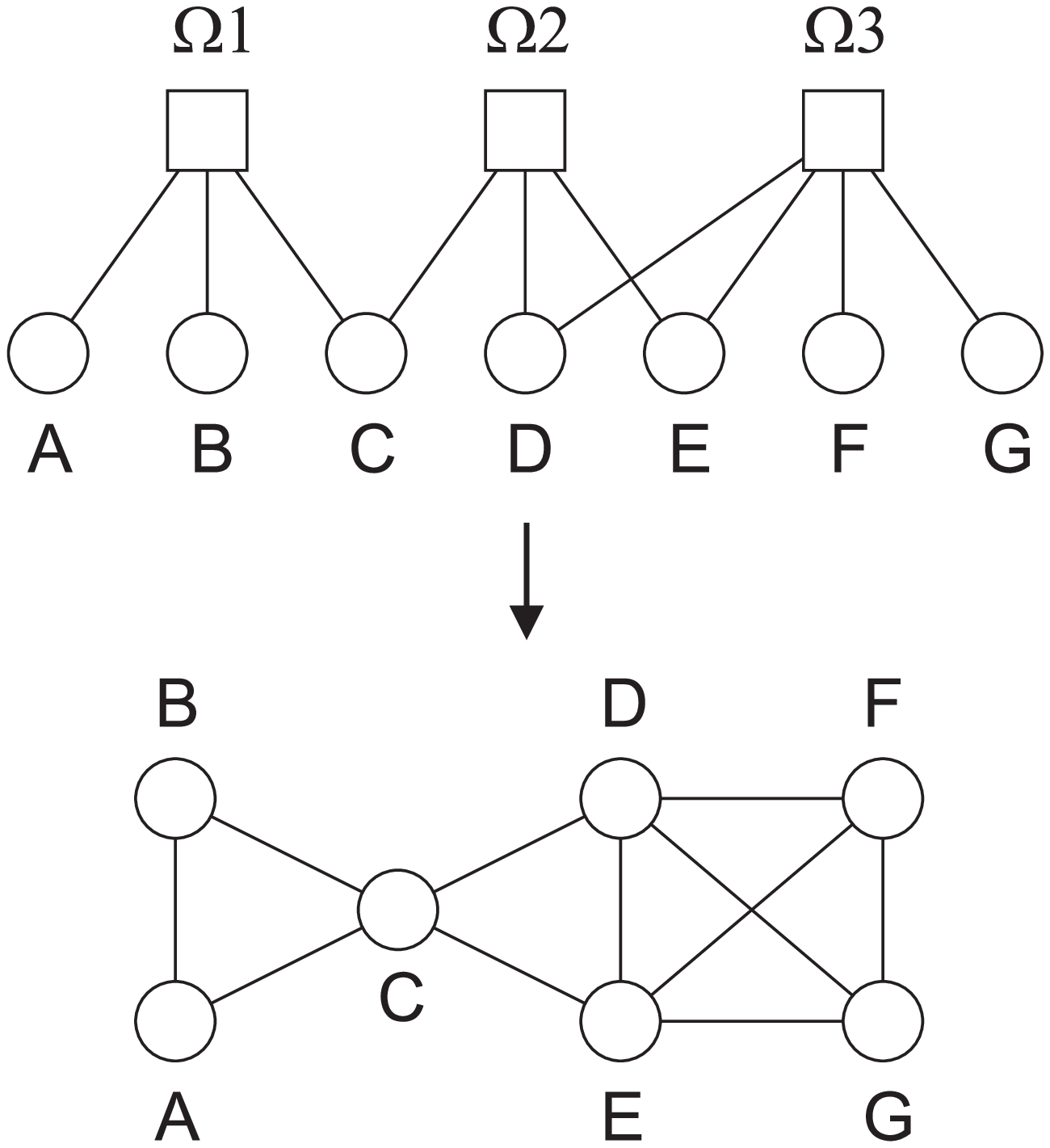}
&
\includegraphics[width=0.3\textwidth]{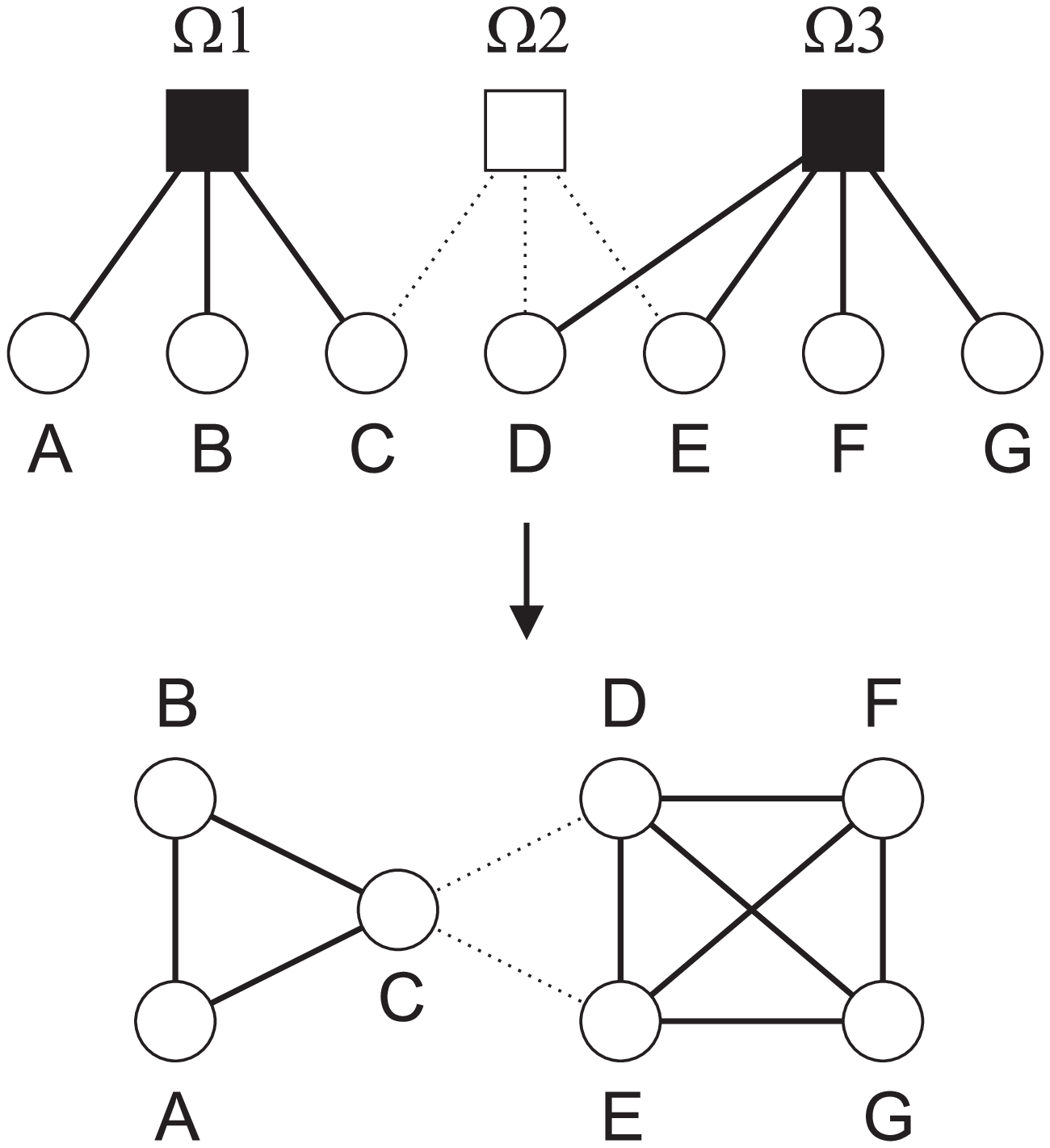}
\\ [0.4cm]
\mbox{(a)} & \mbox{(b)}
\end{array}$
\end{center}
 
\caption{ {\bf (a)} Explanation of the unipartite projection.  Given a
bipartite (or `two-mode') affiliation graph, a new network is defined
on the set of actors, where two actors are connected if they belong to
one or more contexts together in the association graph.  In the above
example, there are seven actors (A--G) and three groups
($\Omega1$--$\Omega3$).  Observe three overlapping cliques in the
one-mode projection (ABC, CDE, and DEFG) corresponding to the three
interaction contexts.  It is possible to differentiate between
different levels of intensity of links in the unipartite projection by
assigning a weight to each context and calculating a summary weight
for each connected pair of actors.  However, for the points we wish to
make here it is sufficient to use the simple undirected graph
representation; that is, to be able to tell if any two actors are
connected or not, neglecting the `strength' of connection.  {\bf (b)}
Boundary Specification Problem for relations.  Suppose that we fail to
include interaction context $\Omega2$ in the above example.  That may
have a drastic effect on the observed properties of the one-mode
network, e.g. it may become disconnected, etc. }

\label{fig:unipro}
\end{figure}

Consider attendance at social events, e.g. Davis's Southern Women
\cite{DaGa41, WaFa94}, or multiple affiliations, e.g. interlocking
boards of directors in American companies~\cite{DaGr97}, or different
interaction contexts (high school students attending classes together
vs going to the movies vs playing sports, and so forth).  Each event,
affiliation or context serves as an opportunity to create, maintain,
or exercise (manipulate) group and interpersonal ties.  The above
examples can be represented by a bipartite graph \cite{Wi82}, in which
one class of vertices represents events, and the second class is
actors.\footnote{  Given the conceptual similarity of affiliation
networks, social event attendance and multiple interaction contexts,
in the discussion that follows we will take the liberty of using the
terms ``events'', ``contexts'' or ``affiliations'' interchangeably,
unless specifically mentioned otherwise. } If an actor participates in
an event, there is an edge drawn between the respective vertices.  To
focus on the class of actors, we perform an operation that is called
unipartite projection, i.e. transformation of a two-mode
``affiliation'' graph into a one-mode network that captures multiple
social relations between the actors (Fig.~\ref{fig:unipro}). One-mode
projections necessarily consist of many overlapping
cliques.\footnote{ Note that a dyad is a clique of size two. } Every
such clique refers to one or several affiliations or interaction
contexts.  In the bipartite framework an affiliation tie is added to
the network if an actor has participated in the given context.
However, correlated contexts are somewhat redundant, in the sense that
they contain much the same information about social structure.  For
example, take a group of coworkers spending a weekend at a picnic
organized by their firm together with their spouses and children.  The
relationships at work and at the picnic may well be different but
daily experience leads us to expect that people who are good
colleagues in the work setting will be likely to socialize with each
other in a semi-formal setting as well.\footnote{ This phenomenon
involves a set of interesting hypotheses which are outside the scope
of this paper but well deserve to be a focus of a separate research
project.  Do people tend to bring their acquaintances from one
interaction context to another? If so, then under what circumstances
does this happen?  In particular, how does the probability of triadic
closure, that is, probability that two friends, A and B, of some
person C, will become friends themselves, depend on the number and
intensity of shared social contexts with C?}

The network approach has traditionally sought to separate different
relational contexts for the sake of analytical tractability.  A
textbook definition of a social network \cite{WaFa94} assumes a
discrete set of actors linked together by a discrete set of relations.
At the interpersonal level, social actors are almost always discrete,
but difficulties arise when we try to disentangle interpersonal
relations such as friendship, help, advice-giving, authority, esteem,
influence, and so on.  It is difficult to devise a classification
scheme that is exhaustive, describes mutually exclusive relations and
has identical meaning to every participating actor.  Multiple
relations are often correlated (e.g. Sampson's data in
\citealp{WhBo76}), that is, people tend to be friends with people that
they like, esteem and can ask for advice, etc.; however, as we have
pointed out, a micro-social mechanism that leads to this correlation
is an open research problem.

Despite the complex structure of interpersonal relations or maybe as a
consequence of it, the resulting pattern of connections is often
perceived as a one-mode network: an overlap of multiple relations,
which perhaps guarantees some protection against misinterpretation of
questionnaire items by respondents or missing important interaction
contexts by researchers, and which is certainly easier to represent
and analyze.  One-mode networks have been studied extensively in the
recent years with a number of important analytic results obtained
\cite{AlJe00, BaAl99, CaNe00, CoEr00, CoEr01, NeSt01, WaSt98}.
However, this line of research has focused on simple models for the
network (e.g. randomly mixed with respect to vertex degree), which are
unlikely to hold in most real situations where both structural and
attribute-based processes are important \cite{GiNe02, WaDo02, Wh92}.
We therefore propose that the multicontextual model of a social
network (generated by a bipartite graph) has certain advantages over
the models based on simple random graphs.  Formulated in a suitable
manner, it is analytically tractable \cite{NeSt01, WaDo02} and by
definition takes care of certain properties observed in empirical
social networks that are not easily reproducible with simple random
graphs (such as high clustering).\footnote{ Some interesting questions
that are related to networks with multiple affiliations or multiple
interaction contexts are the following.  How do network properties
change if new interaction contexts emerge spontaneously?  How should
imputation strategies depend on whether actors create new affiliations
in a competitive or cooperative environment?  Having defined a social
network with several interaction contexts, what is the minimal number
of contexts (the core subset) such that structural characteristics are
robust?  These and related questions will be explored in future
research by analyzing empirical network data and building simulation
models.}

\subsection{An example: forensic data}

While data collection quality in analysis of conventional social
relationships (such as `friendship' or `advice' networks) may be
improved by appropriate research design and cooperation on part of the
participants, the situation in criminal investigation is exacerbated
by the unfortunate fact that criminals seldom cooperate with
law--enforcement agencies.  Not infrequently, they engage in
conspiracy in order to conceal their identities and the structure of
criminal organization.

\begin{figure} 
\begin{center}
\includegraphics[width=\textwidth, bb=-1 14 671 303, clip]{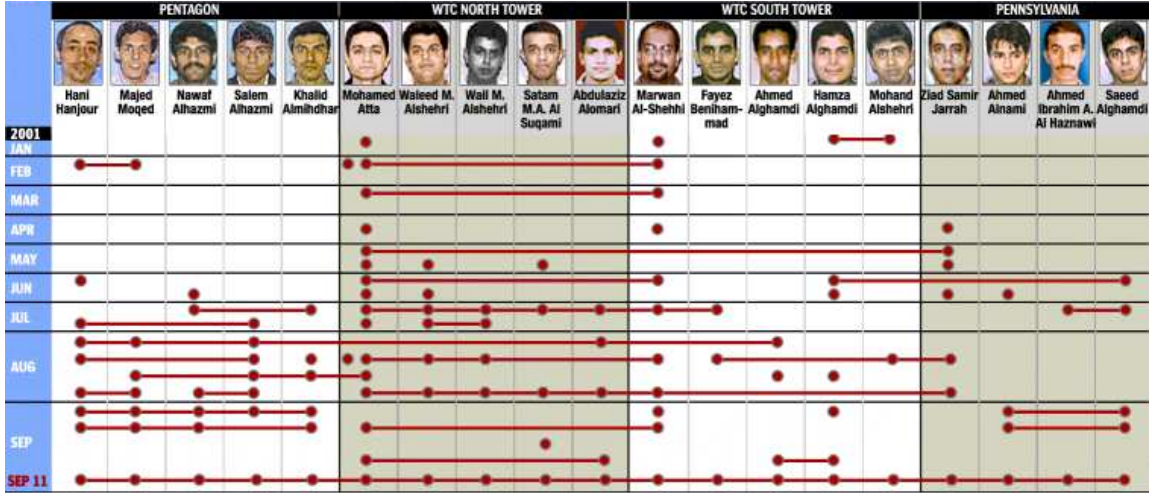}
\end{center}
\caption{\protect The group of September 11 hijackers as an example of
relational BSP.  The chart is reproduced from \citet{WaPost}.  Columns
refer to primary suspects (the hijackers), and dots connected by
horizontal lines represent incriminating contexts, such as: shared an
apartment with another primary suspect, registered for gym membership
with other primary suspects, bought tickets using the same credit
card, etc.  Finally, the latent structure of the criminal network
becomes manifest as all actors participate in the September 11 terror
attacks.  This kind of data naturally maps out as a bipartite graph
where actors are linked by way of interacting in various incriminating
contexts.  Early in the investigation, primary suspects appear to be
linked through a small subset of contexts.  Interaction contexts in a
secret organization are difficult to define and observe for obvious
reasons.  The question is, how many contexts are needed to reconstruct
the structure of the criminal organization with some certainty? }

\label{fig:hj}
\end{figure}

Since investigators typically proceed by expanding ego-networks of
several main suspects, the key actors may be omitted due to ignored or
unknown interaction contexts.  Actors with false or multiple
identities also introduce errors into the structural representation of
the criminal group.  A plausible conjecture is that links may be
easier to uncover once we know the primary suspects (via
surveillance). However, since we expand the circle of suspects by
traversing interactions in certain contexts, missing links are of
great importance, too.

As the result of conspiracy, some meetings, telephone conversations or
email exchanges may not be recorded.  The consequences are two-fold:
first, investigators may be missing certain connections between actors
in the main pool of suspects; second, since those connections lead to
other potential suspects, truncated ties effectively hinder the course
of investigation.\footnote{ It is a single connected component that
investigators seek to obtain. If the unipartite projection of a
criminal network consists of several disconnected components it
probably means that available evidence is not sufficient to conclude
that all actors belong to one criminal group. } We interpret this type
of missing data as the result of incriminating interaction contexts
left outside the scope of analysis.

We suggest that it is natural to represent intelligence data as a
bipartite graph, where suspects are linked to each other through
participation in common actions that we call incriminating interaction
contexts (Fig. \ref{fig:hj}). A single-mode actors network is in fact
a unipartite projection of the intelligence database onto the set of
suspects.  A unipartite projection by definition implies multiple
overlapping cliques.\footnote{ Actions performed by individual actors
are important pieces of evidence that draw attention to these
individuals (call them principal suspects).  Once principal suspects
are known, investigators may proceed with mapping the structure of
criminal network by monitoring actors involved in certain contexts
with the principal suspects (contextual ego-network expansion --
snowballing on the bipartite graph). } Every clique in a network of
criminal organization refers to one incriminating context.  It
therefore follows from the bipartite framework that missing links
usually do not occur alone: they are missing groups of links
corresponding to missed interaction contexts.

Having emphasized the primacy of boundary specification problem in
social network analysis, we now turn to more specific manifestations
of missing data, namely non-response and design effects.\footnote{ The
causes of non-response are outside the scope of this paper.}

\subsection{Non-response effects}
\label{non-response}

The non-response effect in networks with multiple interaction contexts
(modeled as bipartite graphs) is quite different from the same effect
in single-mode (unipartite) networks.  In a survey of an affiliation
network, actors are asked to report groups to which they belong.
Suppose that we have no other sources of information about
affiliations.  If any one actor fails to respond, all his affiliations
are lost and the resulting missing data pattern becomes equivalent to
the Boundary Specification Problem for actors which we model as
stochastic omission of some fraction of actors from the network.

If however the survey asks actors to name peers with whom they
interact (that is, ignoring the multiplexity of ties), then the
non-response effect can be balanced out by reciprocal nominations
\cite{StRi92}.  Suppose actor A did not fill in the network
questionnaire.  Yet those of A's interactants who participated in the
survey must have reported their interactions with A.  Intuitively, one
would expect that if the number of non-respondents is small relative
to the size of the network, and the researcher does not require all
nominations to be reciprocated (as a crude validity check), then the
amount of missing data caused by non-response should be small if not
negligible.\footnote{ Consider a single-mode social network and retain
links that are reported by a) at least one actor; b) both actors only
(the reciprocated subset of nominations).  In this paper we assume the
first mechanism and treat the simplest case of actors not responding
at random, but it would be interesting to consider situations with a)
actors not responding with probability proportional to actor's degree
(call it ``the load effect''); or b) actors not responding with
probability inversely proportional to degree (``the periphery
effect'').  }

\begin{figure} 
\begin{center}
\includegraphics[width=0.25\textwidth]{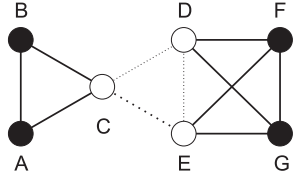}
\end{center}
 \caption{Non-response in network surveys.  Suppose that actors $C$,
$D$ and $E$ did not report their links.  However, nominations made by
actors $A$, $B$, $F$ and $G$ help reconstruct the structure of
interactions to a large extent, with a decrease in average degree less
than 15\%.  Compare with the Boundary Specification example (Figure
\protect \ref{fig:bspa}), in which a single missing node caused a 25\%
deviation in the mean degree. }
\label{fig:nre}
\end{figure}

\subsection{Fixed choice designs}

\begin{figure} 
\begin{center}
$\begin{array}{c@{\hspace{0.7in}}c}
\hspace{-0.1in}\includegraphics[width=0.3\textwidth]{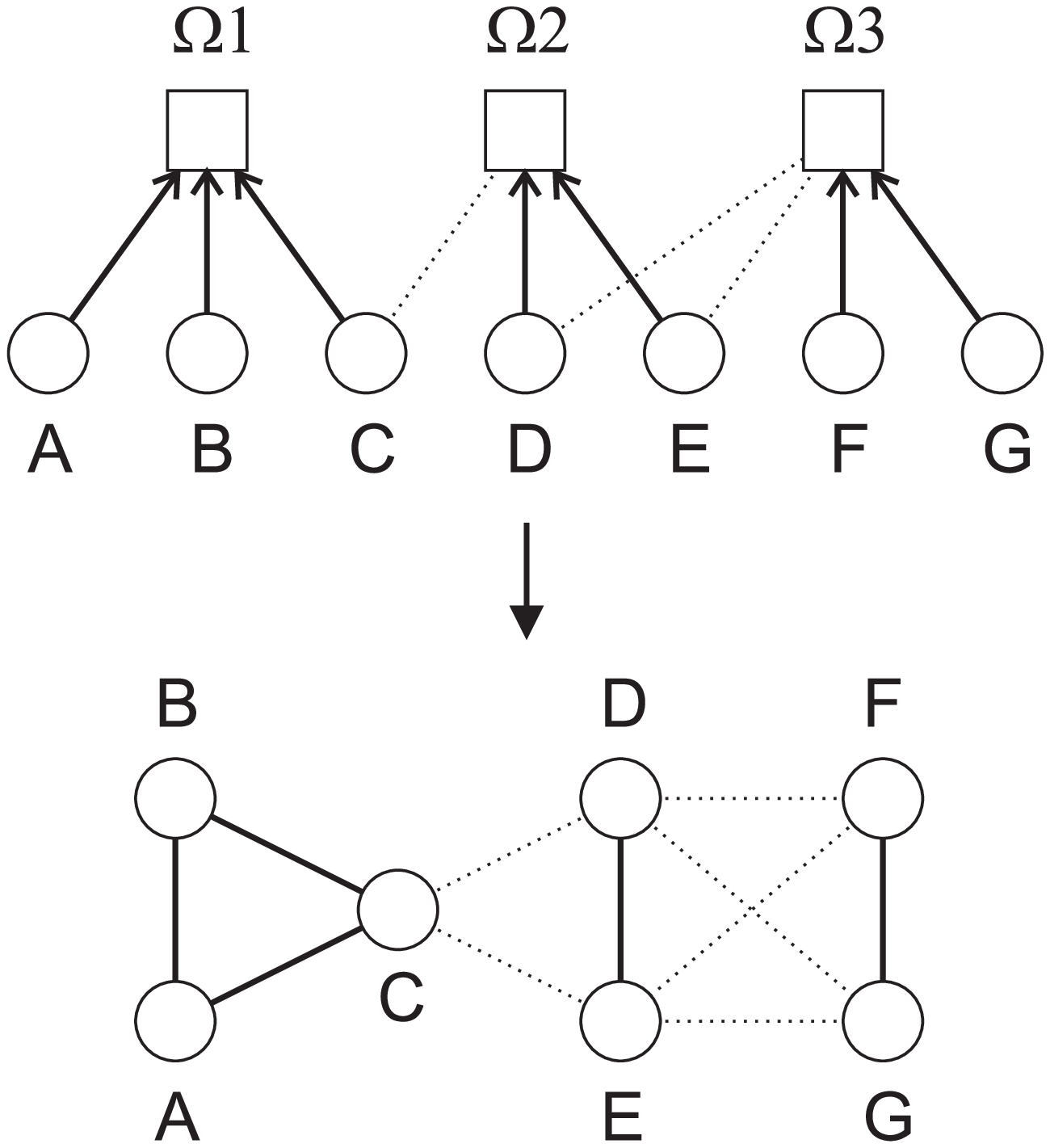}
&
\includegraphics[width=0.25\textwidth, bb= 127 200 456 503, clip]{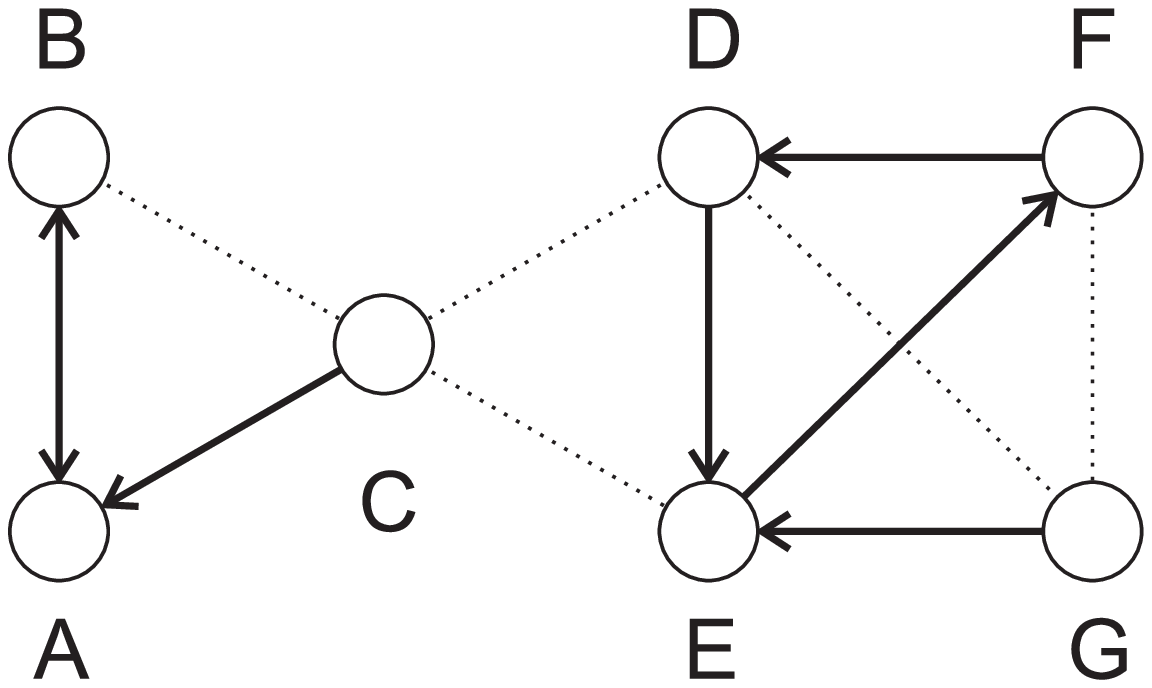}
 \\ [0.4cm]
\mbox{(a)} & \mbox{(b)}
\end{array}$
\end{center}
 \caption{Illustration of a fixed choice design.  {\bf (a)} Bipartite
case: each actor nominates up to a fixed number $K$ from his
affiliations.  Nominations are shown as arrows.  {\bf (b)} One-mode
case: each actor nominates up to a fixed number $X$ from his list of
acquaintances. In the hypothetical example pictured above $K=X=1$.
Note that there is only one reciprocated nomination (between actors $A$
and $B$). }
\label{fig:db}
\end{figure}

Another bugbear of network statistics is right-censoring by vertex
degree (also known as ``fixed choice effect'' \cite{HoLe73}).  This
missing data mechanism is often present in network surveys.  Suppose
that actor A belongs to $k$ groups whereby he is connected to $x$
other actors (Fig. \ref{fig:db}a).  In the unipartite case, the actor
is requested to nominate up to $X$ persons from his list of $x$
interactants, e.g. ``$X$ best friends'' (Fig. \ref{fig:db}b).  If the
cutoff is greater than or equal to the actual number of friends ($X\ge
x$), we assume that all $x$ links between A and his friends are
included in the dataset. If $X<x$, actor A must omit $x-X$ links, but
some of those might still be reported by A's friends who are requested
to make their nominations likewise.  Thus some ties from the original
network will be reported by both interactants (reciprocated
nominations), some by only one partner (non-reciprocated nominations),
and yet some will not be reported (censored links).  It is left to the
discretion of the researcher whether to include non-reciprocated links
which may be qualitatively different from reciprocated ones (e.g.,
good friends vs casual acquaintances).  Fixed choice nominations can
easily lead to a non-random missing data pattern.  For instance,
certain actors may possess some great personal qualities and hence
would be present on the ``best friends'' lists of many other actors.
That is, popular individuals who have more contacts may be more likely
to be nominated by their contacts~\cite{Fe91, Ne03}.  What effect will
this have on the structural properties of the truncated graph?

Generally speaking, selecting randomly from one's list of friends does
not generate a random sample of edges in the graph.  The effect may be
different depending on whether the network is mixed disassortatively
or assortatively by degree~\cite{Ne02, Ne02a, VaMo02}: in the first
case, vertices with high degrees tend to be matched with vertices with
less connections and therefore more censored connections are likely to
be restored using reciprocal nominations.  This is an example of how
the network structure may interact with missing data mechanisms.  


We simulate the fixed choice effect in the following two situations.
First, we consider the bipartite case, i.e. networks with multiple
interaction contexts or affiliations.  We assume that actors are
requested to report up to $K$ groups to which they belong.  We perform
sensitivity analyses for a number of properties of the unipartite
projection as we vary the affiliation cutoff $K$.


Secondly, we simulate a network survey in which actors nominate each
other directly.  To do this we analyze single-mode networks
(i.e. unipartite projections of affiliation graphs) and keep links
that are reported by a) at least one actor; b) both actors only.  For
the sake of simplicity we make the assumption that actors report peers
randomly from their interactant lists.

\section{Data and statistics of interest}
\label{Data}

\subsection{Network-level statistics}
\label{Data:Stats}
As we wish to investigate how topological properties of the network
are affected by the presence of missing vertices or edges, we measure
the following graph-level properties of the unipartite projection onto
actors: mean vertex degree $z$ (average number of interactants per
actor), which characterizes network connectivity;
clustering $C$, that is, the probability that any two vertices with a
mutual neighbor are themselves connected\footnote{ There are several
ways to measure clustering \cite{Wa99, NeSt01, MaSn02}. We adopt the
following definition of clustering coefficient: $C = {3 N_\triangle /
N_3}$, where $N_\triangle$ is the number of triangles on the graph and
$N_3$ is the number of connected triples of vertices.  This definition
is more representative of average clustering in cases when vertex
degree distribution is skewed \cite{NeSt01}.}; assortativity $r$,
which is simply the Pearson correlation coefficient of the degrees at
endpoints of an edge \cite{Ne02, Ne02a}; fractional size of the
largest connected component $S$; and average path length in the
largest component $\ell$.  We accept that the effect of missing data
on parameter $Q$ is tolerable if the relative error
$\epsilon=\frac{|q-q_0|}{q_0}\le 10\%$, where $q$ is an estimate from
a model with missing data and $q_0$ is the respective ``true'' value
calculated from all available data.



\subsection{Data}
\label{Data:Data}
We follow previous work in treating collaboration and affiliation
graphs as proxies of multicontextual social networks \cite{DaMi99,
Mi96, Ne01}.  We illustrate the problem of missing data in networks
using the example of the scientific collaboration graph containing
authors and papers from the Condensed Matter section (``cond-mat'') of
the Los Alamos E-print Archive from 1995 through 1999 \cite{Ne01} as
well as random bipartite graphs. The properties of the dataset are
summarized in Table \ref{tbl:summary}.

\begin{figure} 
\begin{center}
\includegraphics[width=0.9\textwidth, bb=52 242 569 504, clip]{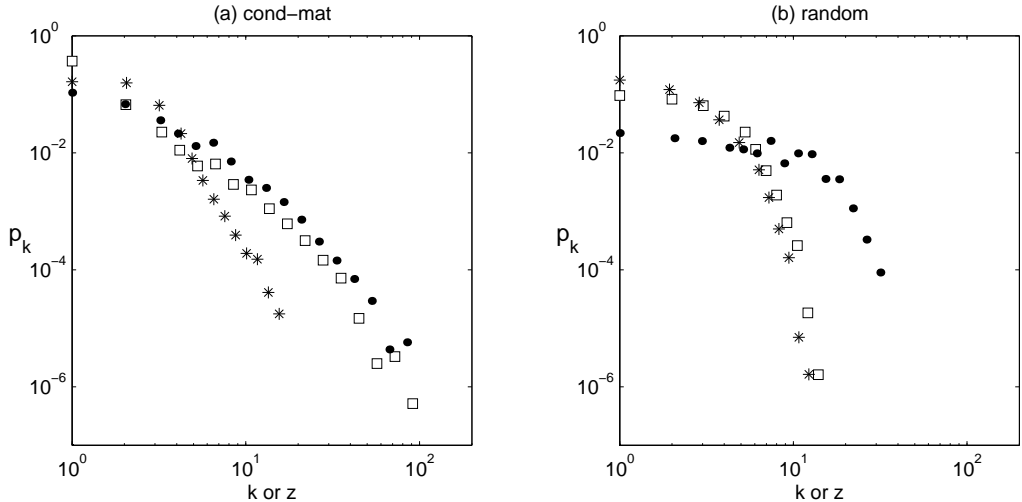}
\end{center}
\caption{ Distributions of vertex degree in the Condensed Matter
collaboration graph (a) and in the comparison random network (b).
{\bf Squares:} number of papers per author; {\bf stars:} number of
authors per paper; {\bf dots:} number of collaborators per author.
The data have been logarithmically binned. }
\label{fig:degreedist}
\end{figure}

We compare the collaboration graph to an ensemble of 100 random
bipartite graphs with the same number of vertices and edges,
i.e. fixing the number of actors $N=16726$, number of groups
$M=22016$, mean degree $\mu=3.50$ for actors and $\nu=2.66$ for
groups\footnote{ Actually, we need to fix only three parameters since
$\mu N = \nu M$. } (Fig. \ref{fig:degreedist}b).  The degree sequences
are not fixed and so they have a Poisson distribution \cite{Bo01,
NeSt01}.  In the Condensed Matter collaboration network, both the
distribution of the number of authors per paper and the distribution
of papers per author are considerably skewed to the left relative to
the random model (Fig. \ref{fig:degreedist}a).  The distribution of
vertex degree in the one-mode coauthor network (i.e.  the number of
co-authors) resembles a power-law with exponential cutoff near $k=100$
(Fig \ref{fig:degreedist}a, dots) while the same distribution in a
random graph exhibits the characteristic bimodal shape \cite{NeSt01}
with a clear cutoff in the tail (Fig. \ref{fig:degreedist}b).  In the
unipartite projection of a random bipartite graph there are many
vertices with a medium connectivity while very few vertices with a
very large number of coauthors.  The values of mean degree in the
one-mode projection are $z=5.69$ for the cond-mat graph and $z=9.31$
for its random counterpart, which indicates a strongly non-random
allocation of authors over papers in the Condensed Matter
collaboration network.  In both cases $z \gg 1$, which guarantees the
existence of the giant connected component \cite{Bo01}.

\begin{table}
\setlength{\tabcolsep}{6pt}
\begin{center}
\begin{threeparttable}
 \caption{Properties of the network dataset.}
\label{tbl:summary}
\begin{tabular}{lccc}
\hline
Quantity & notation & {\tt cond-mat} & random \tnote{a} \\ [0.5ex]
\hline 
Number of authors & $N$ & 16726 & 16726 \\
Number of papers & $M$ & 22016  & 22016\\
Mean papers per author & $\mu$ & 3.50 & 3.50 \\
Mean authors per paper & $\nu$ & 2.66 & 2.66\\ 
Assortativity (degree correlation) & $r_B$  & -0.18 & -0.054(4) \\ [0.5ex]   
Unipartite projection (collaborators):\,& &  \\
\hspace{0.1in} Mean degree & $z$ &5.69 & 9.31(3)\\
\hspace{0.1in} Degree variance & $V$ & 41.2 & 33.9(6)\\
\hspace{0.1in} Clustering & $C$ & 0.36 & 0.223(1)\\
\hspace{0.1in} Assortativity & $r_U$ & 0.18 & 0.071(5) \\
\hspace{0.1in} Number of components & $N_C$  & 1188 & 652(18)  \\
\hspace{0.1in} Size of largest component & $S_L$ & 13861 & 16064(18) \\
\hspace{0.1in} Mean path in largest component & $\ell_L$  & 6.63 & 4.728(8) \\ 
[0.5ex]
\hline
\end{tabular}

\begin{tablenotes}
\item [a] {A random bipartite graph of the same size and mean degree
as the original network. Numbers in parentheses are standard
deviations on the least significant figures calculated in an ensemble
of 100 such graphs.}
\end{tablenotes}

\end{threeparttable}
\end{center}
\end{table}

As seen from Table \ref{tbl:summary} the bipartite form of the
Condensed Matter collaboration graph is disassortative ($r_B=-0.18$)
whereas its one-mode projection is assortative ($r_U=0.18$).  This
implies that authors who work in smaller collaborations publish more
papers on average; also, physicists with many collaborators tend to
work with those of the same ilk; and similarly, physicists with a few
coauthors who are, incidentally, most prolific ones, tend to
collaborate with each other.
\footnote{\label{ft:groupsize}Additional
simulations (not shown here) indicate that the presence of
heavy-tailed group size distribution in a bipartite graph may cause
assortativity in its one-mode projection onto actors.  This lead us to
suggest that assortativity of the one-mode Physics collaboration graph
might be to some extent an artifact of the skewed distribution of
collaboration sizes.}  In addition to providing curious insights into
the mode of scientific production in Condensed Matter Physics,
assortativity has important implications for network robustness \cite{
BoPa03, Ne02, Ne02a, VaMo02}.
A characteristic feature of assortatively mixed ($r_U>0$) networks is
the so-called core group consisting of interconnected high-degree
vertices. The core group provides exponentially many distinct pathways
to connect vertices of smaller degrees.  From an epidemiology point of
view, the core forms a reservoir that is capable of sustaining a
disease outbreak even though the overall network density is too low
for an epidemic to occur.  The good news, however, is that an outbreak
in assortatively mixed networks is likely to be confined to a smaller
subset of the vertices.  Disassortative networks are particularly
susceptible to targeted attacks on high-degree vertices due to the
fact that the latter provide much of the global network connectivity
\cite{Ne03}.

Although a random graph is technically neutral (i.e. has zero
assortativity), it may acquire some disassortativity as a finite-size
effect, e.g. from the constraint forbidding multiple edges between two
vertices \cite{MaSn02, Ne03}.  In a similar fashion, random bipartite
graphs exhibit disassortative mixing if the number of groups differs
from the number of actors.  This follows from the definition of a
bipartite graph (no edges connect vertices of the same class) and the
requirement that no actor belongs to the same group twice.  The
ensemble of random bipartite graphs simulated here exhibit small but
significant disassortativity ($r_B=-0.054\pm 0.004$) while the
corresponding one-mode networks are assortatively mixed by degree
($r_U=-0.071 \pm 0.005$).

It is important to keep in mind that clustering, assortativity (or
generally, the mixing pattern) and degree distribution are not
independent.  In particular, disassortative mixing in simple graphs
may cause a decrease in clustering by suppressing connections between
high degree vertices in favor of vertices of lower degree, thus
reducing the number of triads in the network \cite{MaSn02, Ne03}.


\subsection{Algorithms}

The outline of the simulation algorithm is as follows: (1) take a real
social network or a corresponding ensemble of random graphs; assume
that network data is complete; (2) remove a fraction of entities to
simulate different sources of error; and (3) measure network
properties and compare to the ``true'' values (from the complete
network).  As has been described, we model several missing data
mechanisms.  Table \ref{tbl:algorithms} summarizes our simulation
models.

\begin{table}
\setlength{\tabcolsep}{6pt}
\begin{center}
\begin{threeparttable}
\caption{Simulation algorithms for sensitivity analysis.}
\label{tbl:algorithms}
\begin{tabular}{lp{6cm}p{6.2cm}}
\hline Label & Problem & Model \tnote{a}
\tabularnewline [0.5ex]
\hline
BSPC & Boundary Specification Problem for Contexts & Remove a fraction
of contexts at random \tabularnewline [0.5ex]
BSPA & Boundary Specification Problem for Actors & Remove a fraction
of actors at random \tabularnewline [0.5ex]
NRE & Non-response Effect & Remove links within subgraph induced by a
specified fraction of actors \tabularnewline [0.5ex]
FCC & Fixed choice (contexts) & Apply censoring by degree to actors
\tabularnewline [0.5ex]
FCA & Fixed choice (actors) & Create unipartite projection; apply
censoring by degree; keep non-reciprocated links \tabularnewline
[0.5ex]
FCR & Fixed choice (actors), reciprocated nominations only & Create
unipartite projection; apply censoring by degree; keep only
reciprocated links \tabularnewline [0.5ex] \hline
\end{tabular}

\begin{tablenotes}
\item[a] {We measure properties of the unipartite projection in all
models.}
\end{tablenotes}
\end{threeparttable}
\end{center}
\end{table}

\section{Results and discussion}
\label{Results}

\subsection{Comparison of Boundary Specification and Non-Response Effects}

\begin{figure} 
\begin{center}
\includegraphics[width=0.9\textwidth, bb=72 252 548 468, clip]{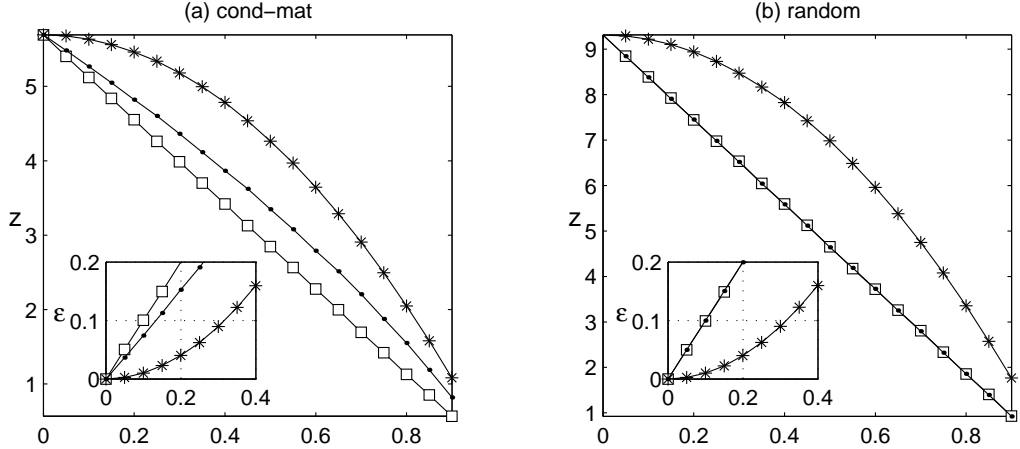}
\end{center}
\caption{ Sensitivity of mean vertex degree in the unipartite
projection $z$ to different missing data mechanisms: (a) in the
Condensed Matter graph; (b) in a bipartite random graph.  {\bf Dots:}
boundary specification (non-inclusion) effect for interaction contexts
(BSPC); the horizontal axis corresponds to the fraction of papers
missing from the database.  {\bf Squares:} non-inclusion effect for
actors (BSPA) with the $x$-axis corresponding to the fraction of
authors missing from the database.  Note that in panel (b) dots
overlap with squares.  {\bf Stars:} simulation of survey non-response
among authors (NRE); vertices are assumed non-responding at random.
The $x$-axis indicates the fraction of non-respondents.  {\bf Insets:}
relative error $\epsilon=|z-z_0|/z_0$, where $z_0$ is the true value.
Each data point is an average over 50 iterations.  Lines connecting
datapoints are a guide for the eye only.  }
\label{fig:me}
\end{figure}

\begin{figure} 
\begin{center}
\includegraphics[width=0.9\textwidth, bb=64 252 548 468, clip]{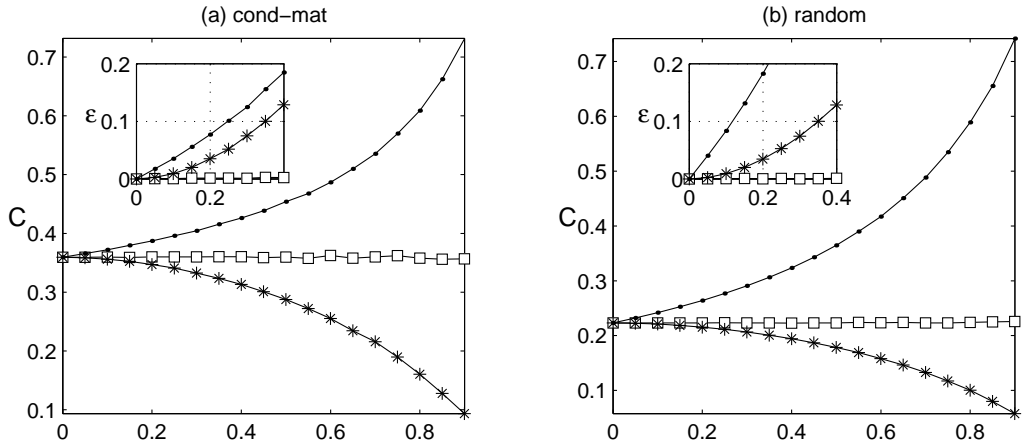}
\end{center}
\caption{ Sensitivity of clustering $C$ in the unipartite projection:
omission of interaction contexts (dots); omission of actors
(squares); survey non-response (stars).}
\label{fig:cm}
\end{figure}

\begin{figure} 
\begin{center}
\includegraphics[width=0.9\textwidth, bb=60 252 548 468, clip]{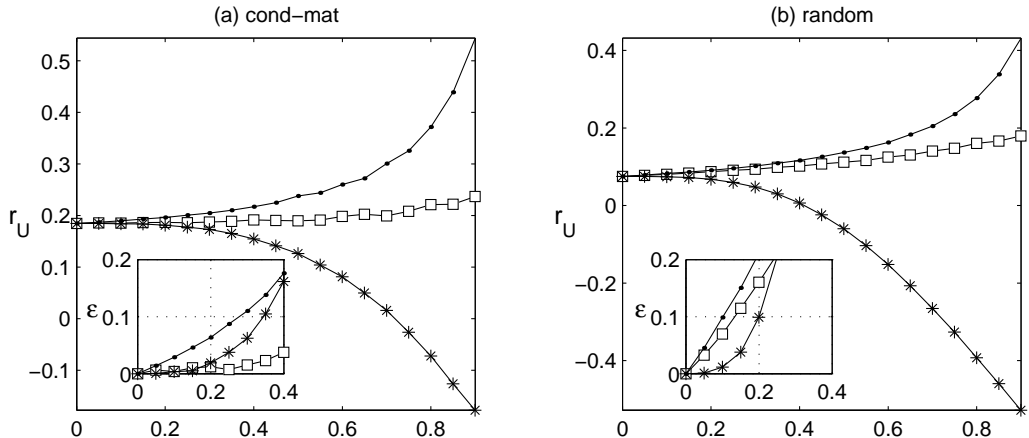}
\end{center}
\caption{ Sensitivity of degree assortativity coefficient $r_U$ in the
unipartite projection: omission of interaction contexts (dots);
omission of actors (squares); survey non-response (stars).}
\label{fig:mx}
\end{figure}

\begin{figure} 
\begin{center}
\includegraphics[width=0.9\textwidth, bb=64 252 548 468, clip]{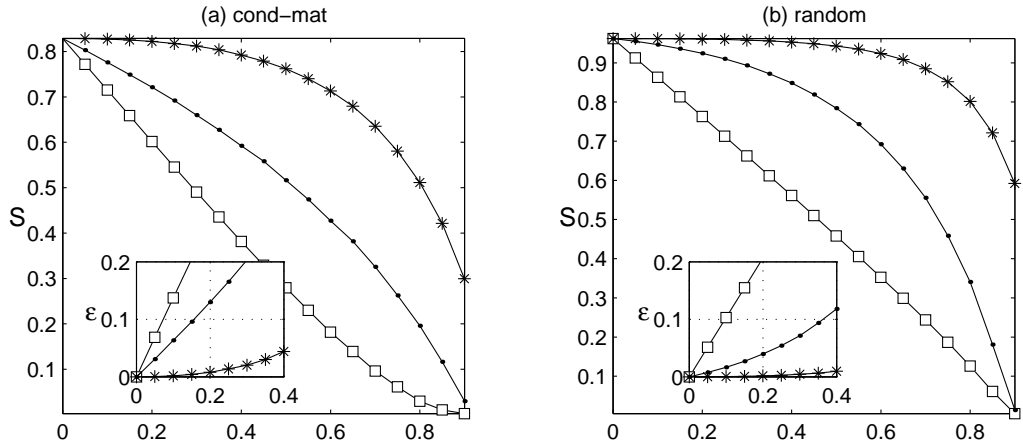}
\end{center}
\caption{ Relative size of the largest connected component in the
unipartite projection: omission of interaction contexts (solid
dots); omission of actors (squares); survey non-response
(stars).}
\label{fig:sl}
\end{figure}

\begin{figure} 
\begin{center}
\includegraphics[width=0.9\textwidth, bb=68 252 548 468, clip]{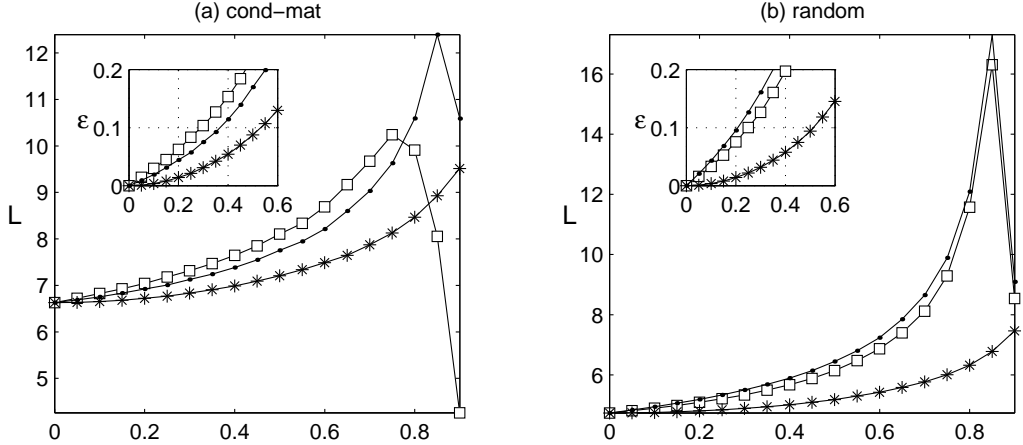}
\end{center}
\caption{ Mean path length in the largest component of the unipartite
projection: omission of interaction contexts (dots); omission of
actors (squares); survey non-response (stars).  Note the drop in path
length corresponding to the lost of connectivity as the network
becomes fragmented and the largest component becomes increasingly
small. }
\label{fig:pl}
\end{figure}

The results of the simulations for the Condensed Matter collaboration
graph and for comparable random bipartite networks are plotted on
Figs. \ref{fig:me}, \ref{fig:cm}--\ref{fig:pl}.  The proportion of
missing data increases from left to right and at the leftmost point we
assume that all information about the network is available.  We model
the Boundary Specification Problem for Contexts (BSPC) by
randomly removing vertices of the corresponding class (``papers'')
from the network. The Boundary Specification Problem for Actors
(BSPA) is modeled as random deletion of vertices corresponding
to ``authors'' in the case of collaboration network.  Survey
non-response is different from BSPA in that in the former vertices are
not removed from the network but all edges between randomly assigned
``non-respondents'' are deleted.

{\bf Mean vertex degree.} For a random bipartite graph, the mean
degree in the unipartite projection onto actors decreases linearly
with random removal of actors or groups: $z=\mu\nu(1-\theta)$, where
$\theta$ is a relative number of missing actors or groups,
respectively\footnote{ Here we have made use of the fact that the mean
vertex degree $z=\mu\nu$ in the unipartite projection of random
bipartite graph, which is symmetrical with respect to changes in
either $\mu$ or $\nu$ \cite{NeSt01}.} (observe overlapping curves in
Fig. \ref{fig:me}b).  However, in the one-mode collaboration network
average degree decreases slower in the simulation of BSPC
(Fig. \ref{fig:me}a, dots) than in BSPA (squares).  This behavior
implies non-random allocation of actors (authors) to groups (papers)
and leads us to introduce the notion of ``redundancy'' in group
affiliation.

\begin{figure}
\begin{center}
\vspace{0.2in} 
$\begin{array}{c} 
\includegraphics[width=0.45\textwidth, bb=97 390 681
531, clip]{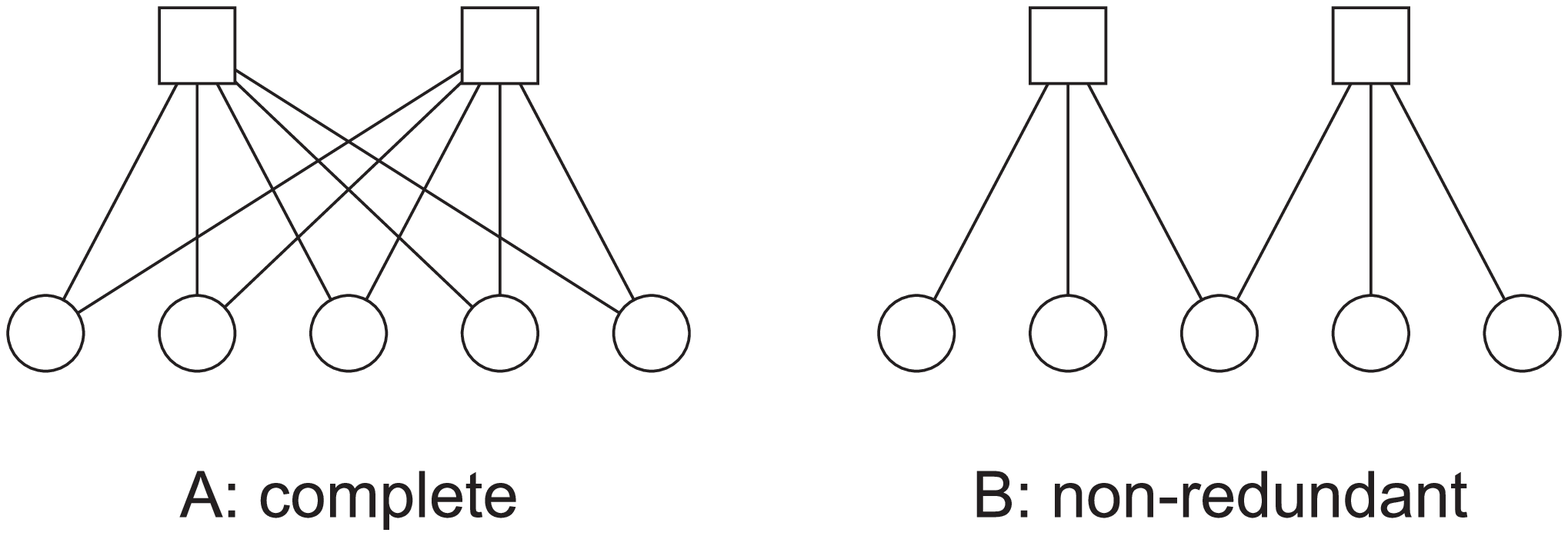} \\ [0.4cm]
\mbox{(a)} \hspace{1.3in} \mbox{(b)} \\
\end{array}$
\end{center}
\caption{Examples of {\bf(a)} complete (maximal redundant); and
{\bf(b)} acyclic (non-redundant) bipartite graphs. }
\label{fig:redun}
\end{figure}

One way to capture the average importance of an interaction context is
to measure what we call the {\em redundancy} of a bipartite graph.  We
define redundancy as
$\beta=\frac{\mu\nu-z}{\mu\nu}=1-\frac{z}{\mu\nu}$, where $\mu$ is
average number of groups per actor, $\nu$ is average size of the
group, and $z$ is actual (observed) mean actor degree in the
unipartite projection onto the set of actors. In a complete bipartite
graph all affiliations but one are redundant in the sense that they
connect actors who are already connected (Fig. \ref{fig:redun}a),
consequently $\beta_C=1-\frac{N-1}{MN}\rightarrow1$ as
$M\rightarrow\infty$ (M is the number of affiliations). At the other
extreme are acyclic bipartite graphs (Fig. \ref{fig:redun}b), in which
if any two actors belong to the same affiliation it is the only
affiliation they share, therefore $z=\mu\nu$ and $\beta_A=0$.
Consider a bipartite graph such that every connected pair of actors
have attended exactly three events together.  The mean degree in the
actors one-mode network will be $z=\mu\nu/3$, and redundancy therefore
is $\beta=1-1/3=2/3$. Redundancy of a random bipartite graph is
expected to be close to zero since $z\approx\mu\nu$, which becomes
exact as the graph size increases \cite{NeSt01}.  In general, high
redundancy implies that as new interaction contexts emerge, they will
likely link already connected actors.  Redundancy of the Condensed
Matter collaboration graph is
$\beta=1-5.69/(3.50\times2.66)\approx0.38$, which means that if the
collaboration sizes were sharply peaked around the mean, then about
forty percent of collaborations could be omitted without any
significant change in the structure of unipartite projection.
However, this is not exactly the case here (Fig. \ref{fig:me}a)
because the group size distribution is quite skewed (Fig.
\ref{fig:degreedist}a).  There are certain important collaborations
that serve as ``hubs'' that stitch together local groups of coauthors,
which may increase the sensitivity of this network to BSPC.
Also recall that the degree correlation coefficient in the original
bipartite network is $r_B=-0.18$, implying that on average authors who
work in smaller collaborations tend to be more productive (this fact
may reflect the nature of the dataset and its limited time frame; see
\citealp{Ne01}).

As could be expected, due to counting in non-reciprocated nominations,
the non-response effect is somewhat less severe than BSP and may be
tolerated for response rates of 70\% and better where the relative
error is less than 10\% (Fig. \ref{fig:me}, insets).

{\bf Clustering.}  Random omission of actors (Fig. \ref{fig:cm},
squares) appears to have no effect on clustering in the unipartite
projection. This result could be expected since all clustering is
engendered via joint membership in groups, whose pattern is unaffected
by random deletion of actors.  It is intuitively plausible that
interaction contexts are responsible for the resulting clustering and
mixing pattern in the bipartite model of a social network.
Fig. \ref{fig:cm} (dots) implies that omission of contexts (BSPC)
results in increased clustering.  As has been mentioned above, each
interaction context or group in a bipartite graph corresponds to a
clique in the one-mode network of actors.  If redundancy of the
bipartite graph is sufficiently high, these cliques tend to overlap.
As more interaction contexts are removed, cliques in the one-mode
network disconnect from each other thus effectively reducing the
number of connected triples of vertices $N_3$ while keeping the number
of triads $N_\triangle$ high.  This causes the clustering coefficient
$C = {3N_\triangle / N_3}$ to grow.

On the contrary, non-response (Fig. \ref{fig:cm}, stars) results in
lower clustering.  Since missing links under non-response are the ones
that connect non-responding nodes and otherwise network connectivity
is not affected, this mechanism opens up triples faster than producing
dyads or isolates, and therefore the clustering coefficient is
decreasing.

The relative deterioration rate (Fig. \ref{fig:cm}b, inset) depends on
the ``true'' value of clustering.  For one-mode networks generated
from random graphs with Poisson degree distributions, clustering
coefficient changes as $C(\theta)=1/(1+\mu(1-\theta))$ in the case of
BSPC, and $C(\theta)$ is fairly close to $\theta/(1+\mu(1-\theta))$
under non-response, where $\theta$ denotes the fraction of missing
groups or non-responding vertices, respectively.  The first result
follows trivially from the formula $C=1/(1+\mu)$, derived by
\citet{NeSt01}; the second is our conjecture based on simulations.

It seems plausible that BSPC and non-response may compensate each
other under some fortunate circumstances, yet separately they
drastically affect the estimate of clustering coefficient and inflate
the measurement error.  Ironically, eliminating one source of error
but not the other could severely impair the estimate of clustering
in the network!

{\bf Assortativity.} The simulation results plotted on
Fig. \ref{fig:mx} show that, as in the case of clustering, BSPC
increases degree-to-degree correlation in the unipartite projection
while non-response causes it to diminish, and ultimately leads to a
disassortative mixing pattern.  We should emphasize these facts as
they increase the uncertainty about the estimates of clustering and
assortativity in networks with unknown missing data patterns.

It has been shown that unipartite networks that are assortatively
mixed by degree are more robust to removal of vertices than
disassortative or neutral networks \cite{Ne02a}.  Several social
networks, including the one-mode collaboration graph analyzed in this
paper have been found to be assortatively mixed.  In such networks,
the assortative core can form a reservoir that will sustain the
disease even in the absence of epidemic in the network at large
(Section \ref{Data:Data}).  As an application to epidemics control,
these findings suggest a rather grim conclusion that social networks
would sustain epidemic outbreaks whereas disease prevention strategies
based on vaccination of high-contact individuals are doomed to fail.

Observe, however, that one tends to overestimate the mixing
coefficient in networks with multiple interaction contexts as a
consequence of the Boundary Specification Problem for Contexts
(Fig. \ref{fig:mx}, dots) and, to a lesser extent, BSP for Actors.
Therefore complete social networks may actually possess less
assortativity than they appear to have, provided that researchers take
measures to minimize non-response.  This finding may turn out to be an
important factor in cost-benefit analyses of disease prevention
strategies that are based on empirical network data.



{\bf Size of the largest connected component.}  As can be seen from
Fig. \ref{fig:sl}, the collaboration network is quite robust to survey
non-response (stars): good estimates can be obtained with response
rates of 70\% and better (50\% for random graphs with similar
parameters).  On the other hand, omission of actors (squares) leads to
immediate and severe deterioration of the network connectivity.  The
effect of missing interaction contexts (dots) is somewhere in-between.
From the modeling point of view, non-inclusion of actors (as well as
actor non-response with required reciprocation, for that matter) is
equivalent to the so-called ``node failures'' analyzed in several
recent studies of computer networks \cite{AlJe00, CaNe00, CoEr00,
CoEr01, VaMo02}.  This line of literature has focused on the effects
that random failures or intentional attacks on Internet routers might
have on the global connectivity properties of the Internet, such as
the size of the largest connected component.  In particular, it has
been shown that for random breakdowns, networks whose degree
distribution is approximated by a power-law remain essentially
connected even for very large breakdown rates \cite{CoEr00}.  It has
been also demonstrated under quite general assumptions that
disassortativity increases network fragility as it works against the
process of formation of the giant component; on the other hand,
assortative correlations make graph robust to random damage
\cite{VaMo02}.  However, our simulation results do not fully agree
with these notions.  The one-mode coauthorship network is
assortatively mixed and has a heavy-tailed degree distribution, while
the projection of a random bipartite graph has near zero assortativity
and quickly decaying degree distribution (Fig. \ref{fig:degreedist} a
and b respectively, dots).  Yet under BSPA the size of the largest
component decreases faster in the one-mode collaboration network
(compare Fig. \ref{fig:sl}a and Fig. \ref{fig:sl}b, squares).

To separate possible effects of mixing pattern and degree
distribution, we have run simulations with bipartite networks obtained
by randomly rewiring the collaboration graph.  These networks have the
same degree sequences as the original bipartite graph but zero
assortativity coefficient.  The rewired networks behave very similarly
to random graphs with Poisson degree distribution.  An important
difference, however, is that random removal of actors initially leads
to a faster decrease in the size of the giant component $S_L$, but for
large removal rates $S_L$ approaches zero size continuously in a
rewired network (not shown here), while both random graph and the
original collaboration network exhibit a discontinuity (easily seen in
the plot of average path length, Fig. \ref{fig:pl}).  We conclude that
a rewired version of the collaboration graph is more resilient to BSPA
than the original, despite its lack of assortativity.  Hence,
assortativity alone does not necessarily imply network robustness,
contrary to previous assertions, and may have substantially different
implications for networks engendered via joint membership in groups or
interaction contexts.  The compound effect of the mixing pattern and
degree sequences in such networks therefore deserves a further
investigation.



{\bf Mean path length in the largest connected component.}  As may be
seen from Fig. \ref{fig:pl}, BSPA and BSPC have a similar effect on
the average path length.  Path length diverges when mean vertex degree
becomes less than unity.  Due to the skewed degree distribution of the
Condensed Matter collaboration network BSPA has a stronger impact on
mean degree than BSPC, and consequently, the phase transition
(breakdown of the largest component into many small ones) occurs at
$\theta\approx0.75$ for BSPA and $\theta\approx0.9$ for BSPC.  The
effects of missing data mechanisms on the mean path length may be
tolerated (i.e. relative error not exceeding 10\%) for amounts of
missing data up to 20\% in case of BSPA or BSPC, and for response
rates of 50\% and better in case of actor non-response.

\subsection{Degree censoring (fixed choice effect)}

\begin{figure}  
\begin{center}
\includegraphics[width=0.9\textwidth, bb=68 252 548 468, clip]{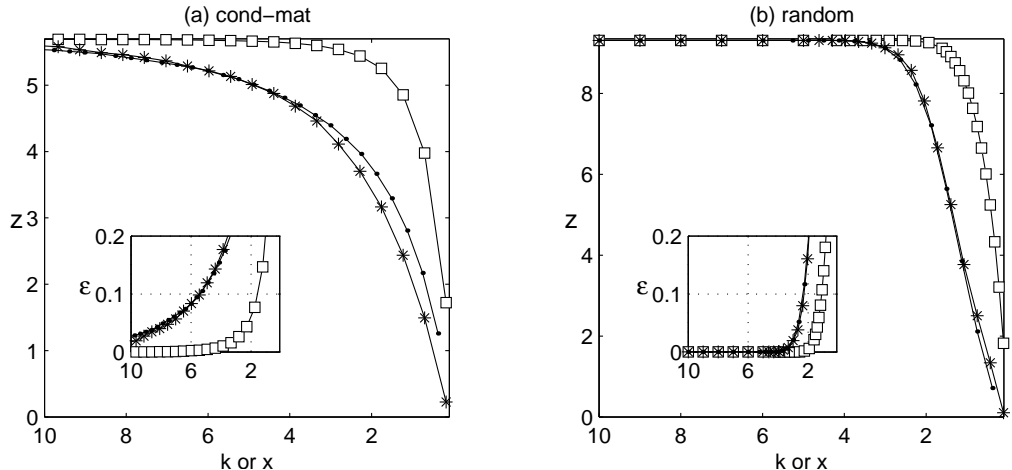}
\end{center}
\caption{Fixed choice effect on the mean degree of the unipartite
projection $z$ in the Condensed Matter collaboration graph (a) and a
comparable random graph (b).  {\bf Dots:} censoring collaborations.
The question asked of each author would be to ``nominate'' up to $K$
papers coauthored by him.  The horizontal axis represents the relative
degree cutoff $k=K/\mu$, where $\mu=3.5$ is the mean number of
affiliations per actor.  Note that the amount of missing data
increases as we lower the threshold value.  For example, $k=5$ means
that the actual cutoff is $K=5\mu$, five times the mean actor degree
in the bipartite network.  {\bf Squares:} censoring coauthors, no
reciprocation required.  The question asked of each author would be to
nominate up to $X$ coauthors.  The horizontal axis represents relative
degree cutoff $x=X/z$ in units of $z$, the mean number of
collaborators per author, where (a) $z=5.69$ in the Physics
collaboration graph and (b) $z=9.31$ in a random network. {\bf Stars:}
only reciprocated nominations, relative cutoff $x=X/z$ in units of
$z$.  {\bf Insets:} relative error $\epsilon=|z-z_0|/z_0$, where $z_0$
is the true value.
Each data point is an average over 50 iterations.  Lines connecting
datapoints are a guide for the eye only.  }
\label{fig:dbme}
\end{figure}

We consider the impact of fixed-choice questionnaire design
(right-censoring by vertex degree) on network properties in the
following three cases: (1) we record up to $K$ interaction contexts
out of average $\mu$ for every actor; (2) each actor nominates up to
$X$ out of average $z$ interaction partners; the link is present if
either one or both members of a dyad report it; (3) same as previous,
but every dyadic link must be reported by both partners.  Varying the
cutoff values $K$ and $X$, we have explored how these missing data
mechanisms affect the unipartite social network under assumption of
random nominations.  Sensitivity curves for the mean vertex degree are
shown on Fig. \ref{fig:dbme}.  The results for other statistics
discussed in the previous sections are qualitatively similar to the
corresponding BSP/non-response effects up to the direction of error
(see Tables \ref{tbl:fractions} and \ref{tbl:cutoffs} for details).

It appears that degree censoring has a much more severe effect on the
Condensed Matter collaboration graph (left plot) than on a random
bipartite network with the same parameters $N$, $M$ and $\mu$ (right
plot).  In a random graph, a fixed choice of $K=k\mu$ interaction
contexts (collaborations) or reciprocated nomination of $X=xz$
partners practically does not affect mean degree $z$ as long as
relative cutoffs $k>3$ or $x>3$.  In the collaboration graph, however,
mean degree departs from its true value as soon as the relative cutoff
$k$ or $x$ becomes less than $15$.  As a consequence, this impairs
estimates of such network properties as the number of components, size
of the largest component and geodesics length (not shown).  The
effects of degree censoring on network properties are quantified in
Table \ref{tbl:cutoffs}, where we report approximate minimal cutoff
values such that parameter estimates are within $\pm10\%$ around their
respective true values.  It is noteworthy that fixed choice errors are
virtually non-existent in random graphs for relative cutoff values $k$
or $x\gtrsim 2$.  On the contrary, the real collaboration network
appears to be very sensitive to degree bound effects.

While there may be a number of different mechanisms at work, it is
likely that this difference in behavior is a joint effect of the
non-random mixing and skewed degree distributions observed in the
Condensed Matter collaboration graph.  Censoring by degree has little
effect on the random graph because its degree variance is quite small,
i.e. it is rather sharply peaked around the mean.  Therefore, when we
cut edges in excess to, say, $2\mu$ or $2z$ in a random graph, the
number of actually removed links is negligible.  On the other hand,
the distribution of papers by authors and the distribution of the
number of collaborators in the one-mode network both have a heavy tail
(Fig. \ref{fig:degreedist}), i.e. there is a considerable fraction of
vertices with degrees greater than twice the average value.  If the
one-mode network is mixed assortatively by degree as in the case of
the Condensed Matter graph, then degree censoring will likely
eliminate most connections within the network core and quickly break
down the giant component.  Additional computer experiments (not shown)
with a randomly rewired version of the cond-mat network, which has the
same degree distribution but zero mixing, support this explanation.
Whereas skewed actor degree distribution alone may have a limited
impact on the robustness of network statistics with respect to the
fixed choice effects, when present together with assortative mixing,
it makes the network increasingly more sensitive.  We would like to
stress that one-mode projections of bipartite graphs, assortativity
may arise as a structural artifact of a skewed group size distribution
(see footnote \ref{ft:groupsize}), rather than being a substantive
property of some network process.  Hence it is important when doing
empirical research that possible fixed choice effects be carefully
examined if there are reasons to think that the network under study
has been engendered by a multicontextual affiliation graph.

\begin{table}
\setlength{\tabcolsep}{2pt}
\begin{center}
\begin{threeparttable}
 \caption{ Approximate tolerable fractional amount of missing
data\tnote{a} and direction of deviation\tnote{b} for boundary
specification and non-response effects}
\label{tbl:fractions}
\begin{tabular}{lcccc}
\hline Property of one-mode network & Symbol & BSPC \tnote{c} & BSPA \tnote{d}
& NRE \tnote{e}\\ [0.5ex]
\hline 
Mean degree & $z$ & 
 0.14 (0.1)\tnote{f} $\downarrow$ & 0.1 (0.1) $\downarrow $  & 0.3 (0.3) $\downarrow$\\ [0.5ex]
Clustering & $C$ & 
0.25 (0.1) $\uparrow$ & n.a.\tnote{g}  & 0.35 (0.35) $\downarrow$  \\ [0.5ex]
Degree correlation & $r_U$ & 
0.3 (0.1) $\uparrow$ & n.a. (0.15) $\uparrow$ & 0.35 (0.2) $\downarrow$  \\ [0.5ex]
Size of largest component & $S_L$ & 
0.15 (0.35) $\downarrow$ & 0.08 (0.1) $\downarrow$ & n.a. \\[0.5ex]
Mean path in largest component & $\ell_L$ & 
0.4 (0.2) $\uparrow$ & 0.3 (0.25) $\uparrow$ & 0.5 $\uparrow $\\ [0.5ex] \hline
\end{tabular}

\begin{tablenotes}
\item[a] {Missing data is tolerable if it causes relative error not
exceeding 10\%, i.e. $\epsilon=|\frac{q-q_0}{q_0}|\le 0.1$, where $q$
is an estimate from a model with missing data and $q_0$ is the value
calculated from complete data. }
\item[b] {We use $\uparrow$ or $\downarrow$ to indicate the
direction of departure of the estimate from the true value (up or
down, respectively) for a small amount of missing data such that the
network is kept above the percolation threshold, i.e.  mean vertex
degree $z>1$. }
\item[c] {Boundary specification for interaction contexts or
affiliations}
\item[d] {Boundary specification for Actors (missing actors)} 
\item[e] {Non-response, reciprocated nominations are not required}
\item[f] {Numbers in parentheses are results for an ensemble of 100
random bipartite graphs with the same number of vertices and edges.}
\item[g] {Very slow change: less than 10\% error for 50\% of missing
data.}
\end{tablenotes}

\end{threeparttable}
\end{center}
\end{table}

\begin{table}
\setlength{\tabcolsep}{3pt}
\begin{center}
\begin{threeparttable}
 
\caption{Approximate minimal tolerable cutoffs\tnote{a} and direction
of deviation\tnote{b} for degree censoring
effects}
\label{tbl:cutoffs}
\begin{tabular}{lcccc}
\hline Property (projection) & Symbol & FCC \tnote{c} & FCA \tnote{d}
& FCR \tnote{e}\\ [0.5ex]

\hline Mean degree & $z$ & $5.5\mu$ (2.5)\tnote{f} $\downarrow$ & $1.5z$ (1) $\downarrow$ &
$5.5z$ (2.5) $\downarrow$\\ [0.5ex]

Clustering & $C$ & $8\mu$ (2.5) $\uparrow$ & $1.5z$ (1) & $6z$ (1.6) \\ [0.5ex]

Degree correlation & $r_U$ & $18\mu$ (3.5) $\uparrow$ & $6z$ (2.5) $\downarrow$ &
$6z$ (2.5) $\downarrow$\\ [0.5ex]

Size of largest component & $S_L$ & $3.5\mu$ (1.2) $\downarrow$ &
$1z$ (0.2) $\downarrow$ & $2z$ (0.7) $\downarrow$ \\ [0.5ex]

Mean path in largest component & $\ell_L$ &
$6.5\mu$ (2) $\uparrow$ & $1.8z$ (0.9) $\uparrow$ & $5z$ (2) $\uparrow$\\ [0.5ex] 

\hline
\end{tabular}

\begin{tablenotes}
\item[a] { The degree cutoff is tolerable if the relative error caused
by censoring $\epsilon=|\frac{q-q_0}{q_0}|\le 10\%$, where $q$ is an
estimate from a model with missing data and $q_0$ is the value
calculated from complete data. }
\item[b] {We use $\uparrow$ or $\downarrow$, where applicable,
to indicate the direction of departure of the estimate from the true
value (up or down, respectively) for a small amount of missing data
such that the network is kept above the percolation threshold, i.e.
mean vertex degree $z>1$. }
\item[c] {Fixed choice of interaction contexts} 
\item[d] {Fixed choice of actors, reciprocation not required} 
\item[e] {Fixed choice of actors, only reciprocated nominations}
\item[f] {Numbers in parentheses are results for an ensemble of 100
random bipartite graphs with the same number of vertices and edges. }
\end{tablenotes}

\end{threeparttable}
\end{center}
\end{table}





\section{Some implications for empirical analysis}
\label{Implications}

In practice it might be difficult to estimate the effects of missing
data and to identify and separate its sources.  Therefore one should
take measures against multiple possible missing data effects.  The
findings reported in this paper are based on a case study and
statistical simulations of random graphs and therefore may not apply
to all social networks.  However, some of the results are quite
general and enable us to offer some guidelines for researchers who
have collected or plan to collect empirical network data, to help them
be aware of potential pitfalls.

Our simulations indicate that three most severe missing data problems
are: (1) boundary specification for interaction contexts (BSPC); (2)
boundary specification for actors (BSPA); (3) fixed choice designs
(usually FCA, i.e. actors nominating up to a certain number of
partners).  Boundary specification can dramatically alter estimates of
network-level statistics, in particular, the assortativity coefficient
and mean degree, even if context redundancy is large.  In a fixed
choice survey design, the errors introduced by missing data are
relatively small up to certain degree cutoff values, which depend on
the vertex degree distribution and mixing pattern; the worst case
being networks with highly skewed degree distributions, which may
produce unreliable statistics, especially in the presence of
assortative mixing.

These results have the following implications.  In studies which
employ a fixed choice design \cite[e.g.][]{BeMo02}, if there are
reasons to expect a heavy tail distribution, it is crucial to choose a
relatively high degree cutoff to minimize the impact of missing data
on network statistics.  Furthermore, if the network is expected to be
assortatively mixed, the fixed choice design might not be appropriate
at all, and it would be better to use an open list questionnaire,
i.e. allowing respondents to nominate as many partners as they deem
relevant.  Alternatively, one may want to first obtain rough estimates
of the mean degree $z^*$ and its standard deviation $\sigma_z^*$ using
a small sample \cite{Gr76} and simply asking with how many actors
from within the network a respondent has interacted during the
specified period of time.  If $\sigma_z^*>>z^*$ then at the step of
collecting full network data one should employ an open list design or
set the cutoff as high as possible.

A similar double-stage strategy might be appropriate, if not always
feasible, for designs based on formal group affiliation to help
minimize the amount of missing data due to the boundary specification
problem.  After the sociometric data is collected inside an
organization, one should calculate the network diameter $D$.  At the
second step, traverse via other relevant interaction contexts for $D$
removes outside the organization (since the longest possible cycle in
the network is $2D$ long).  If the connectivity properties of the
network (i.e. the number of components and average geodesic length) as
well as clustering and assortativity coefficients do not change
significantly, that implies that the organizational network in
question is robust with respect to boundary specification.  In the
example of adolescent sexual network in a high school \cite{BeMo02},
if the above procedure indicated robustness then persons with outside
partners could be modeled as having higher infection probabilities
with the network model otherwise intact.

Finally, for forensic research it seems most important that the
network of suspects is well-connected so that investigators can start
from a few principal actors and ``snowball'' to the rest of suspects.
As we have found that the size of the largest connected component is
very sensitive to the omission of actors, an obvious recommendation
would be to expand surveillance at the early stages in the
investigation.

\section{Conclusions}
\label{Conclusions}

In this paper, we have set out to compare different missing data
mechanisms in social networks with multiple interaction contexts.
Social interactions are modeled as a bipartite graph, consisting of
the set of actors and the set of interaction contexts or affiliations.
The conventional single-mode network of actors is a unipartite
projection of the bipartite graph onto the set of actors.  We have
measured structural properties of this projection while varying the
amount of missing data in the generating bipartite graph by omitting
actors, interaction contexts, or individual interactions.  This paper
has covered several missing data mechanisms; in particular, boundary
specification and fixed choice survey design.  As a proxy of a
multicontextual social network we analyzed the Los Alamos Condensed
Matter collaboration network and an ensemble of random bipartite
graphs with similar parameters.

Since we have analyzed a specific empirical case and the corresponding
ensemble of random networks, the findings reported herein may not be
generalizable.  With all due limitations, several results of
particular significance follow from our studies.  First, we found that
assortativity coefficient is overestimated via omission of interaction
contexts (affiliations) or fixed choice of affiliations.  On the other
hand, actor non-response or fixed choice of collaborators leads to an
underestimated mixing coefficient and may even cause an assortatively
mixed network to appear as disassortative.  For example, this may
explain why the adolescent romantic network \cite{BeMo02} that was
constructed using fixed choice nominations was found to be neutrally
mixed by degree, in a stark contrast to the majority of known social
networks \cite{Ne02a}.

In a similar fashion, the observed clustering coefficient increases
via omission of interaction contexts or fixed choice thereof, and
decreases with actor non-response.  The clustering coefficient is
unaffected by random omission of actors since all clustering in the
bipartite model of social networks is engendered via interaction
contexts (group affiliation).  The divergent effect of the two missing
data mechanisms obviously results in inflated the measurement error.
It is ironic that by eliminating one source of error (e.g.,
non-response) but not the other (boundary specification effect) one
might actually end up with worse estimates of clustering or
assortativity.

Finally, the confounding effect of mixing pattern and degree
distribution on network robustness under random omission of actors is
found to be different from what is assumed in the current literature.
We have found that under certain circumstances a network assortatively
mixed by vertex degree is less robust to random deletion of vertices
than a comparable neutral network.  As a tentative explanation, we
attribute this peculiar behavior to the detailed structural
composition of the networks that we have focused on; namely, the
presence of multiple overlapping cliques in the one-mode network as a
result of unipartite projection.  Consequently, we would like to
emphasize the importance of further research to better understand the
roles and properties of multiple interaction contexts in the
emergence, evolution, and study of social networks.

The results reported in this paper have been obtained using the method
of numerical simulation.  While this approach is frequently employed
in statistics, it appears underrepresented in network research.
However, we find that it is particularly well-suited for exploratory
analysis of large-scale networks.  Thanks to its power and
flexibility, the method of statistical simulation shows promise as a
useful addition to existing network analysis toolkits.  We hope that
the classification scheme and the systematic exploratory approach that
we have presented will prove useful for further research in the field.



\bibliography{networks}
\addcontentsline{toc}{section}{References} 
\end{document}